\documentclass[12pt]{spieman}  
\usepackage{amsmath,amsfonts,amssymb}
\usepackage{upgreek}
\usepackage{graphicx}
\usepackage{gensymb}
\usepackage{setspace}
\usepackage{tocloft}
\usepackage{lineno}
\usepackage{mathrsfs}

\DeclareMathOperator{\erf}{erf}

\title{Quantum Efficiency Measurements of a Low-Cost, Commercial CMOS Sensor in the Soft X-ray Bandpass}

\author[a, *]{Casey T. DeRoo}
\author[a]{Mallory Weber}
\author[a]{Carson Bantz}
\author[a]{Cole Primeaux}
\author[b]{Hannah Gulick}
\author[c]{Colin M. Packard}
\author[d]{Philip Kaaret}
\affil[a]{University of Iowa, Department of Physics and Astronomy, Iowa City, Iowa, United States}
\affil[b]{California Institute of Technology, Department of Astronomy, Pasadena, CA 91125, United States}
\affil[c]{Naval Research Laboratory, Space Science Division, Washington, DC 20375, United States}
\affil[d]{NASA Marshall Space Flight Center, Huntsville, AL 35812, United States}

\cftpagenumbersoff{figure}
\cftpagenumbersoff{table} 
\begin{document} 
\maketitle

\begin{abstract}
Complementary metal-oxide semiconductor (CMOS) image sensors offer distinct advantages as imaging detectors for space-based X-ray observatories, being low-power, radiation-resistant, and operable at room temperature. Previous characterizations of the low-cost, commercially-available IMX290LLR-C back-illuminated CMOS sensor from Sony Semiconductor Solutions have found Fano-limited X-ray spectral resolutions while operated at room temperature and a radiation tolerance compatible with use in low Earth orbit. Here, we report the quantum efficiency of the IMX290LLR-C as measured across the soft X-ray band (200 - 1000 eV) using Beamline 6.3.2 of the Advanced Light Source at Lawrence Berkeley National Laboratory. We find a quantum efficiency of roughly 10 -- 40\% over this band, and fit these measurements to a quantum efficiency model constrained by these measurements and microscopy of the sensor layers. These measurements and the derived model informs this sensor's application within low-cost X-ray observatories, as well as motivates the development of commercial X-ray CMOS sensors without microlenses and deeper Si depletion layers.
\end{abstract}


{\noindent \footnotesize\textbf{*}Casey T. DeRoo,  \linkable{casey-deroo@uiowa.edu} }

\begin{spacing}{2}   

\section{Introduction}
\label{sect:intro}  

The wide range of electronic circuits that can be produced using complementary metal–oxide–semiconductor (CMOS) processes has made them the backbone of modern semiconductor manufacturing. CMOS active pixel sensors are manufactured \textit{en masse}, with applications in cell phone cameras, security, and production line monitoring. While visible light imaging consistutes the vast majority of applications, CMOS image sensors have also been utilized as soft X-ray (0.2 -- 1.0 keV) detectors in scientific settings, such as synchrotron measurements \cite{Desjardins2019}, X-ray micrograph imaging and inspection \cite{Lane2012,  Hoidn2015, Castoldi2015, AlcaldeBessia2018}, and in astronomy \cite{Miles2019}. 

CMOS image sensors have several advantages over the charged-coupled device (CCD) technologies utilized by major X-ray imaging missions operating in this portion of the X-ray bandpass (e.g., NASA's Chandra X-ray Observatory \cite{Weisskopf2002}, ESA's XMM-Newton \cite{Short1998}). CMOS devices are comparatively low power, realize significantly faster read-out rates, and are more radiation-resistant due to their embedded pixel architecture. Furthermore, CMOS sensors achieve noise performance comparable to CCDs even while operated at room temperature. In contrast, operational temperatures of order -100°C are typical for X-ray CCDs (e.g., \cite{ChandraPOGChap6, Reynolds2023, Grant2024}). These differences are particularly significant for X-ray CubeSats (e.g., HaloSat \cite{Kaaret2019}, BlackCAT \cite{Falcone2025}) or SmallSats (GoSOX \cite{Marshall2026}), which are constrained in budget, mass, and power and subject to radiation damage during their operation in low Earth orbit (LEO).

The back-illuminated (BI) Sony IMX290LLR-C CMOS sensor is designed for visible light applications; however, previous investigations have shown it to be a promising candidate for X-ray detection in space-based observatories. Initial X-ray measurements of this sensor have demonstrated energy resolutions comparable to those achieved by the Chandra ACIS and Suzaku XIS focal plane arrays \cite{Tammes2020}. Follow-on investigations have shown that Fano-limited energy resolutions are achievable by performing a gain calibration on a per-pixel basis \cite{Schneider2024}, and that the sensor showed no change in spectral resolution following a total ionizing dose equivalent to four years in LEO \cite{Roth2022}. 

The detection sensitivity of a space-based X-ray observatory using the Sony IMX290LLR-C depends on the sensor's quantum efficiency (QE). Previously, a Sony IMX290LLR-C sensor was measured to have a QE of 28 $\pm$ 2\% at $E$ = 490.5 eV, but this measurement was limited to a single energy due to the geometry of the experimental configuration.\cite{Packard2024}. In this work, we report on the QE of the IMX290LLR-C CMOS sensor as measured across the soft X-ray bandpass from 200 --  1000 eV. These measurements, along with measurements of the thicknesses of the sensor layers taken via scanning electron microscopy (SEM), are used to construct a QE model from which the sensitivity of X-ray instruments employing this sensor can be estimated. 

\section{Experiment Overview}

Quantum efficiency is the ratio of detected flux to incident flux. As an X-ray source, we utilize Beamline 6.3.2 of the Advanced Light Source (ALS) synchrotron at the Lawrence Berkeley National Laboratory. The incident flux is measured using a dedicated photodiode within Beamline 6.3.2's experiment chamber. The flux detected by the IMX290LLR-C is calculated by operating a customized, flange-mounted camera in integrating mode. Custom \emph{python}-based software is used to store full frames, which are converted from a digitized signal to a detected power through application of a gain calibration. By performing these measurements at a series of soft X-ray energies, this experiment yields an empirical QE as a function of energy.

\subsection{Experimental Setup}
\label{subsec:experimental-setup}
\begin{figure}[t]
\begin{center}
\begin{tabular}{c}
\includegraphics[width=5.5in]{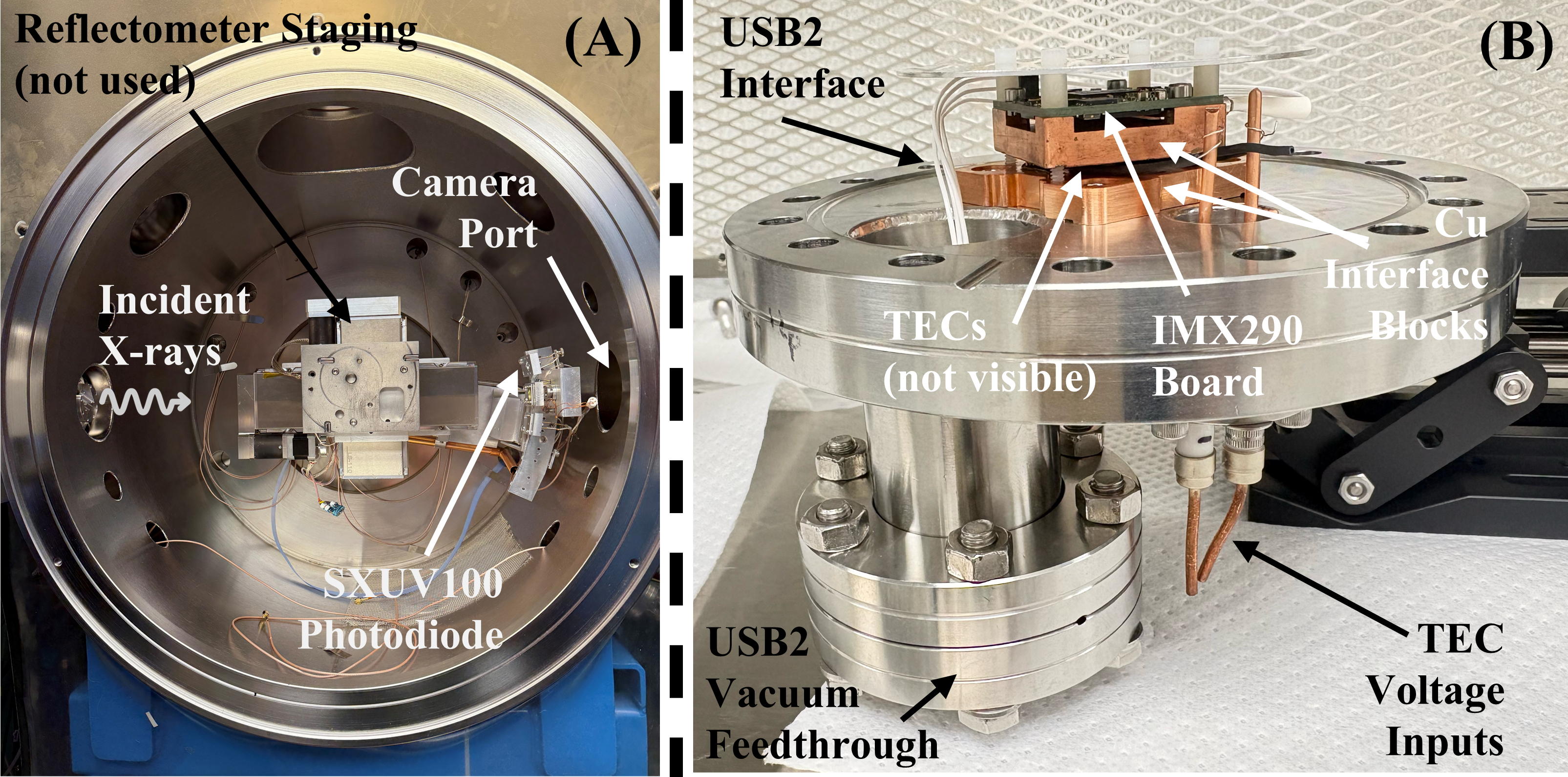}
\end{tabular}
\end{center}
\caption 
{ \label{fig:als-setup}
(A) The reflectometer end station of Beamline 6.3.2 of ALS. X-rays are focused and monochromatized using an optical system upstream, and enter the endstation from the left. These X-rays are either detected using a SXUV100 photodiode to measure the incident flux, or pass through the endstation to the flange-mounted IMX290LLR-C camera through the labeled vacuum port on the right. The reflectometer staging, in the center of the endstation, is designed to facilitate measurements of mounted samples for other types of experiments conducted at Beamline 6.3.2, but is not utilized for this experiment. (B) The interior of the X-ray camera featuring the IMX290LLR-C. The X-ray camera uses TECs sandwiched between two Cu interface plates to maintain a suitable operating temperature while at vacuum. A USB2 vacuum interface powers the camera and transmits image data. A reducer cone (not shown) forms a vacuum enclosure and provides a mechanical interface to the endstation. } 
\end{figure}

Beamline 6.3.2 of ALS \cite{Underwood1996, Underwood1998} is a bending magnet X-ray beamline operating in the energy range of 50 -- 1300 eV. It employs a system of focusing mirrors and variable-line spaced gratings to produce a monochromatic ($E/$$\Delta$$E >$ 7000), well-focused beam on a sample mounted within a reflectometer. X-ray flux is detected using a \hyperlink{https://optodiode.com/wp-content/uploads/2025/02/SXUV100DS.pdf}{SXUV100 photodiode from Opto Diode} that can be translated in and out of the beam using a concentric rotation stage. For this experiment, Beamline 6.3.2's X-ray reflectometer was translated out of the way and the IMX290LLR-C camera was mounted directly in the path of the X-ray beam. Fig. \ref{fig:als-setup}A shows the interior of the 6.3.2 endstation.

The assembled X-ray camera system is shown in Fig. \ref{fig:als-setup}B. The IMX290LLR-C sensor under test is integrated in a \hyperlink{https://en.ids-imaging.com/store/ui-3860le.html}{monochrome UI-3860LE camera from IDS Imaging}. The camera board was removed from its plastic housing. The cover glass of the sensor was mechanically removed in order to make it sensitive in the soft X-ray bandpass. The IMX290LLR-C is equipped with a microlens array mounted atop the active Si; removal of this microlens array was attempted through mechanical scraping. Both removal operations were performed through an external vendor. 

For testing, the camera board was mounted to two copper interface plates, which were in turn secured to a custom 6" CF flange serving as a baseplate. To transfer the heat generated during operation away from the sensor, two \hyperlink{https://www.digikey.com/en/products/detail/marlow-industries-inc/nl1013t-01ac/6159074}{NL1013T-01AC thermoelectric coolers} (TECs) operated at a voltage of 14 Volts and a current of 0.6 Amps installed in series were sandwiched between the copper interface plates. PEEK spacers and screws are used to assemble this interface in order to avoid conducting heat from the hot interface plate to the cold interface plate via the fasteners. A thin, circular Al plate with a cutout matching the dimensions of the IMX290LLR-C (5.6 mm $\times$ 3.1 mm) is mounted along the beam path to prevent the synchrotron beam from irradiating the camera board. Power and data are transmitted over a four-wire USB2 interface. A reducer cone connects to the custom 6" CF on which the camera is mounted, and forms a vacuum housing for the sensor. This cone terminates in a 2.75" CF flange that connects to Beamline 6.3.2's endstation. 

We note that, since obtaining the UI-3860LE, this camera has been depreciated in favor of a variant operating over USB3 \hyperlink{https://en.ids-imaging.com/store/u3-3860le-rev-1-2.html}{(U3-3860LE-M-GL Rev. 1.2)}. As the U3-3860LE-M-GL Rev. 1.2 still employs the IMX290LLR-C sensor, however, we expect identical X-ray QE and spectral performance.




\subsection{Data Collection}
\label{subsec:data_collection}
X-ray measurements were collected during a single campaign spanning Oct. 23\textsuperscript{rd} through Oct. 24\textsuperscript{th}, 2025. X-ray measurements were conducted from 200 eV to 1000 eV in 50 eV steps. Incident flux on Beamline 6.3.2's SXUV100 photodiode was determined by averaging over 150 seconds, while the dark current of the photodiode was measured by translating it out of the beam. The storage ring current of the synchrotron was logged during both the photodiode measurements and measurements with the IMX290LLR-C. The measured storage ring current provides a means to normalize for any temporal variation in the X-ray flux produced by the synchrotron itself between the two data sets; these logs indicate a maximum variation of $<$ 0.2\% in the storage ring current over the entire course of the test campaign.

All measurements with the IMX290LLR-C are performed in native analog-to-digital units (ADU). For each energy, 3000 full frame images were collected. A temperature sensor on the printed circuit board of the camera monitor is used as a record of the device temperature during each frame, and is stored along with each full frame image in a .FITS file \cite{Wells_1981}. At the conclusion of the X-ray test campaign, three sets of 3000 dark frames were taken with the endstation's beam stop in place. 

The exposure time and gain settings of the UI-3860LE camera were selected to enable detection of Beamline 6.3.2's bright X-ray beam without saturation. Beamline 6.3.2 provides count rates of $10^{7}$ -- $10^{11}$ photons/sec. in normal operation \cite{Underwood1996}. These fluxes are orders of magnitude outside those encountered in a typical astronomical setting, where X-ray sensors are operated in photon-counting mode and detect only a handful of X-ray events per frame. To overlap this camera dynamic range with this high intensity source, measurements were performed at 7.4 frames/sec., the maximum frame rate available in 12-bit camera operation in the \emph{python}-based software we developed for our camera. This implies a nominal exposure time of 135 ms for all IMX290LLR-C frames. Similarly, the software gain setting was set to 1 (i.e., the lowest possible). We note that this is different from the settings employed in previous work with the IMX290LLR-C \cite{Tammes2020, Roth2022}, which were optimized for X-ray energy resolution. Finally, we employed a black level offset of 2 ADU for all frames.

\section{X-ray Data Reduction}
\label{sect:data_reduction}
To calculate QE and the associated error, both measurements with the IMX290LLR-C and the photodiode were converted from native units to a detected power. Power from the photodiode is calculated with a vendor-provided responsivity curve. For the IMX290LLR-C, this was accomplished by (1) subtracting a temperature-dependent average dark signal from full-frame images, (2) performing an energy calibration of the detector under experimental conditions using the spectra garnered from single photon events, and (3) using this calibration to deduce the energy in X-rays deposited during a single exposure based on the integrated signal within a full frame. 

\subsection{Temperature Calibration}
\label{subsec:temp_cal}

\begin{figure}
\begin{center}
\begin{tabular}{c}
\includegraphics[width=5.5in]{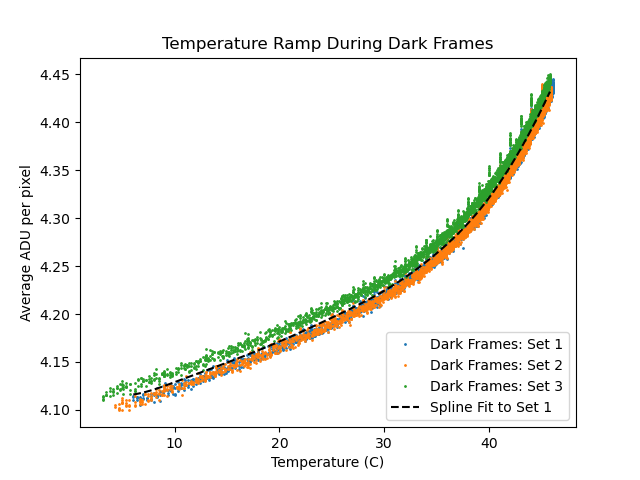}
\end{tabular}
\end{center}
\caption 
{ \label{fig:temp-dependence}
The average detector dark signal per pixel as a function of temperature. Measurements are taken from each of the three sets of 3000 dark frames. From these data, a spline interpolation is computed in order to predict the temperature-dependent background signal in X-ray measurements taken with the the IMX290LLR-C.} 
\end{figure}

While the commercial TECs provided a means to cool the IMX290LLR-C during vacuum operation, they provided insufficient heat dissipation to maintain a steady-state temperature during the operation of this test configuration. Hence, the average temperature varied over the course of the conducted X-ray measurements, increasing from approximately 5$\degree$C to 45$\degree$C during the collection of 3000 data/dark frames. Using the three sets of dark frames and the camera's on-board temperature measurement, the average detector signal per pixel as a function of temperature was tabulated. The average ADU per pixel was calculated by first calculating the raw signal average $\overline{s_{raw}}$ and standard deviation $\sigma_{raw}$. These raw image statistics were used to filter out hot pixels. Any pixels reporting a value greater than $\overline{s_{raw}}$ +  10$\sigma$ were masked, and the average ADU per pixel was computed from the pixels remaining in the given full frame image. A least-squares cubic spline was fit to the three data sets, such that the average dark signal as a function of temperature can be subtracted from the X-ray signal recorded by the IMX290LLR-C. Fig. \ref{fig:temp-dependence} shows these data and the spline fit. We note that the average dark signal changes by $\lesssim$10\% over this relatively broad range of operating temperatures, illustrating one of the advantages of X-ray CMOS technology noted in Sec. \ref{sect:intro}.

\subsection{Energy Calibration}
\label{subsec:energy_cal}

\begin{figure}
\begin{center}
\begin{tabular}{c}
\includegraphics[width=5.5in]{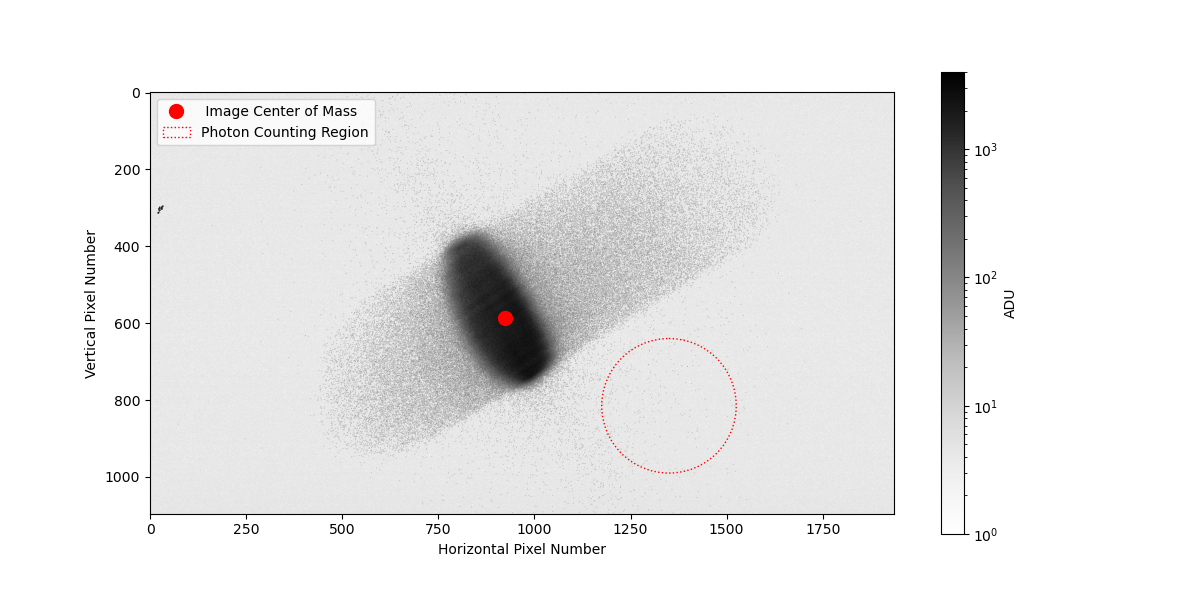}
\end{tabular}
\end{center}
\caption 
{ \label{fig:xray-data}
A full-frame measurement of the X-ray beam using the IMX290LLR-C at an energy $E$ = 600 eV. The pixel size is 2.9 $\upmu$m square, making this image approximately 3.1 mm vertically by 5.6 mm horizontally. Image data are plotted on a log scale. A small region of hot pixels resulting from damage to the sensor's surface is visible at rough coordinates of (50, 300). The center of mass of the image is denoted with a red dot, while an off-beam region used for extracting single photon events is shown as a dashed circle.} 
\end{figure}

A typical X-ray measurement of the X-ray beam of Beamline 6.3.2 by the IMX290LLR-C camera is shown in Fig. \ref{fig:xray-data}. Prior to collecting the data set of 3000 frames, the brightness of the X-ray beam was tuned through insertion of Beamline 6.3.2's transmission filters to yield a maximum pixel intensity of approximately 2000 ADU, i.e., half the maximum data value of 4096 during 12-bit operation, in order to prevent pixel saturation. Visual inspection of data frames at all energies show no evidence of `clipping,' suggesting that the beam is fully captured within the IMX290LLR-C.

To calculate the conversion between the detector's native ADU units and photon energy, we identified a region of interest dominated by single photon events in each data frame and employed photon-counting to produce a spectrum at each X-ray beam energy. The photon counting algorithm developed for this sensor is described in detail elsewhere \cite{Tammes2020, Roth2022}, and is briefly summarized here. Sets of dark frames are used to build a list of hot pixels that are excluded from all measurements, as well as an average dark frame. Each data frame has this average dark frame subtracted, and is then searched for pixels that exceed an pre-defined event threshold. For each such pixel, a surrounding 5 $\times$ 5 grid is searched for pixels that exceed the split pixel threshold and are simply connected to the central pixel. Pixels that exceed these thresholds are considered ``hit pixels,'' and summed to yield a total energy in ADU.  Each such instance is considered a single photon event for which position, frame number, and summed energy are recorded. A region of interest in each of the 3000 data frames is subjected to this procedure, and used to construct an event list for each energy. For all X-ray energies measured, the resultant X-ray event list contains $>$ 250,000 counts.

\begin{figure}
	\begin{center}
		\begin{tabular}{c}
			\includegraphics[width=5.5in]{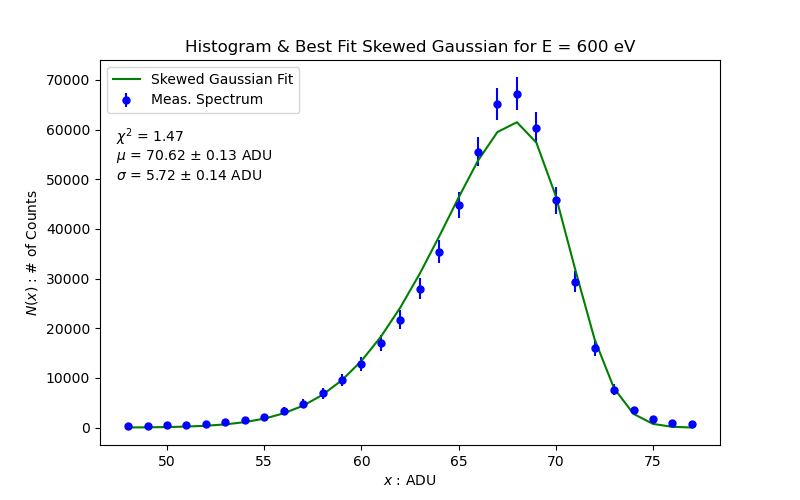}
		\end{tabular}
	\end{center}
	\caption 
	{ \label{fig:xray-spectrum}
		An exemplar X-ray spectrum constructed using the photon-counting algorithm described in Sec. \ref{subsec:energy_cal}. Photon-counted spectra and best-fit models are used only for the purpose of establishing an energy calibration of the IMX290LLR-C under the operating conditions employed at ALS, selected to maximize the dynamic range of the sensor.  A low energy tail resulting from the incomplete counting of split events broadens the full-width half maximum (FWHM) to roughly $100 - 120$ eV over the energy range. } 
\end{figure}

Binning the event list by ADU value produces a spectrum. An example X-ray spectrum produced via this photon counting process is shown in Fig. \ref{fig:xray-spectrum}. The change in software gain to unity in order to maximize the sensor's dynamic range (see Sec. \ref{subsec:data_collection}) requires changing the threshold values used in the photon counting algorithm compared to its previous applications to IMX290LLR-C X-ray data\cite{Tammes2020, Roth2022, Packard2024}). An event threshold of 16 ADU and a split pixel threshold of 6 ADU was utilized for this analysis. We selected these thresholds in order to minimize the uncertainty in the peak centroid position $\mu$, which is used to fit for the sensor gain and therefore an input to the IMX290LLR-C QE estimate. However, these more stringent threshold settings, coupled with a decrease in software gain, makes the employed photon counting algorithm less sensitive in detecting split events. This yields a low-energy tail to the measured spectrum, and degrades the energy resolution by roughly a factor of 2$\times$ over the energy resolution previously achieved with this sensor and photon-counting algorithm\cite{Tammes2020}. We emphasize these changes in operating conditions and the resulting degradation of energy resolution are employed only to enable a QE measurement using the high-intensity synchrotron source. Furthermore, we note that, as QE is a detector property independent of source intensity, the measurement reported here is applicable in photon-counting applications such as X-ray astronomy.

To each extracted spectrum, we fit a skewed Gaussian of the form:

\begin{equation}
	\label{eq:skewed_gaussian}
	N(x) = \frac{A}{\sigma \sqrt{2\pi} } \exp\left(- \frac{(x - \mu)^2}{2\sigma^2} \right) \{1 + \erf\left[\frac{\gamma (x - \mu)}{\sigma\sqrt{2}}\right] \} \, ,
\end{equation}

\noindent where $x$ is the bin number in ADU, $\mu$ is the centroid position in units of ADU, $\sigma$ is the standard deviation of the non-skewed Gaussian distribution, $\gamma$ is related to the skew of the resulting distribution, and $\erf$ is the error function. Least-squares fitting is performed using \hyperlink{https://lmfit.github.io/lmfit-py/index.html}{\emph{lmfit}}, a Python-based curve-fitting package \cite{Newville2026}. Errors in $N(x)$ are assumed to be of the form $\alpha_E \sqrt{N}$, where the factor $\sqrt{N}$ is derived from a standard Poisson treatment and $\alpha_E$ is a scaling factor empirically derived by requiring that $\chi^2_{\nu} \sim 1$ for each energy. X-ray energies $E > $ 350 eV are satisfactorily fit by the applied skewed Gaussian model. The applied event threshold of 16 ADU creates a discontinuity in the low energy tail of the spectrum for energies of 200, 250, and 300 eV; we opt not to utilize the derived ADU centroid from these energies in our energy calibration.

\begin{figure}
	\begin{center}
		\begin{tabular}{c}
			\includegraphics[width=5.5in]{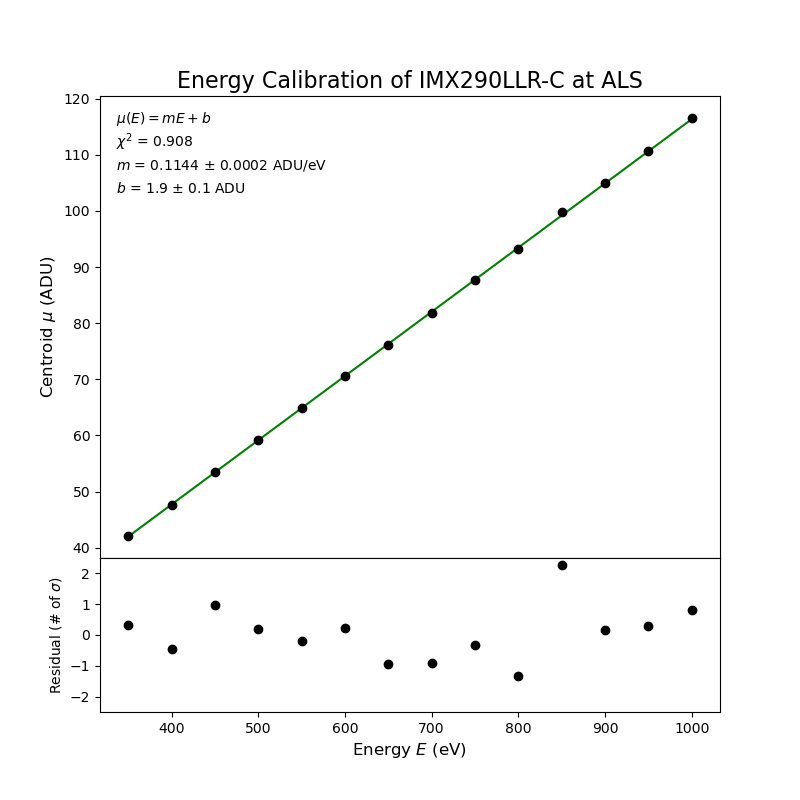}
		\end{tabular}
	\end{center}
	\caption 
	{ \label{fig:gain_fit}
		The derived energy calibration for the IMX290LLR-C under the ALS testing conditions described in Sec. \ref{sect:data_reduction}. Black points indicate the best-fit $\mu$ obtained through spectral fitting at that X-ray energy; error bars in $\mu$ are plotted but smaller than the data marker. A linear model is adequate to describe the data ($\chi^2_{\nu} = 0.908$), yielding an energy calibration of $8.741 \pm 0.015$ eV/ADU. 
		} 
\end{figure}

We use the best-fit centroids $\mu$ as function of X-ray energy to empirically derive the energy calibration of the IMX290LLR-C under these conditions, as seen in Fig. \ref{fig:gain_fit}. Derived fit errors in $\mu(E)$ from each skewed Gaussian fit are used as 1$\sigma$ errors in constructing this energy calibration, and range from 0.1 -- 0.2 ADU. A linear model, $\mu(E) = mE + b$, where $m$ is the slope (units: ADU/eV) and $b$ is the intercept (units: ADU) describes the data well, with a $\chi^2_{\nu} = 0.908$. This finding is in keeping with the high degree of signal linearity previously reported for the IMX290LLR-C\cite{Juneau2026}. The best-fit slope is found to be $m = 0.1144 \pm 0.0002$ ADU/eV while the best-fit intercept is $b = 1.9 \pm 0.1$ ADU. Inverting the best-fit slope yields an energy calibration of $8.741 \pm 0.015$ eV/ADU, equivalent to 2.40 e$^{-}$/ADU given the average hole-pair creation energy of 3.64 eV in Si at room temperature\cite{Scholze1996}. 

\subsection{Direct Beam Power Measured with 6.3.2 Photodiode}
\label{subsec:photodiode}
\begin{figure}
	\begin{center}
		\begin{tabular}{c}
			\includegraphics[width=5.5in]{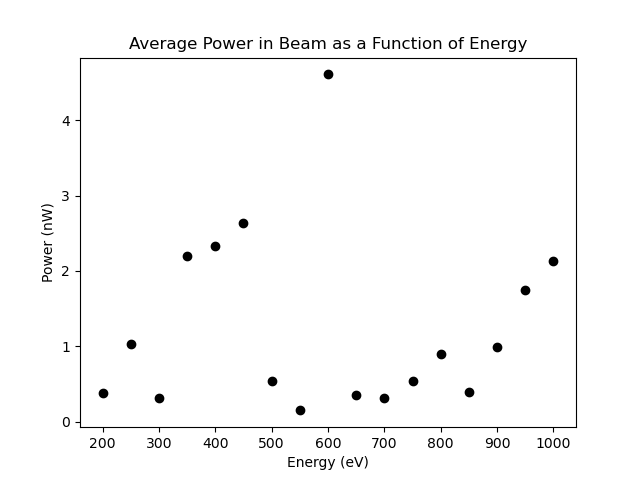}
		\end{tabular}
	\end{center}
	\caption 
	{ \label{fig:photodiode-power}
		Power contained in the direct X-ray beam as a function of energy. Measurement configurations were selected to avoid saturating the detector. The photon rates, of order 10$^7$ photons/sec., represent relatively low fluxes for Beamline 6.3.2.
		} 
\end{figure}

At each energy tested, the power in the X-ray beam was measured using a SXUV100 photodiode. This diode, measuring 10 mm $\times$ 10 mm, is sufficiently large to capture the full extent of the X-ray beam (see Fig. \ref{fig:xray-data}). Average powers at each energy, shown in Fig. \ref{fig:photodiode-power}, were calculated from a time series of 151 independent current measurements. A background current is measured by translating the photodiode out of the X-ray beam and subtracted from the time series; in all cases, the measured current exceeds the background by $\gtrsim$10$\times$. Measurements are normalized based on the current in the ALS storage ring recorded at that time step, which serves as a proxy for any time variations in the X-ray beam intensity during measurement. These background-corrected, normalized currents are then averaged and converted into a power using the responsitivity of the photodiode to yield $P_{Direct}(E)$. As the power in the direct beam is dependent on the configuration of the monochromator and upstream filters selected for a given energy, a smooth dependency as a function of energy is not expected. The range of powers, approximately 0.1 -- 4 nW, represent $\sim$10$^{7}$ photons/sec, and hence cannot measured in the camera's photon-counting mode as noted in Sec. \ref{subsec:data_collection}.

\subsection{Beam Power Detected by IMX290LLR-C}
To extract the power detected by the IMX290LLR-C $P_{IMX290}(E)$, we compute the average total signal for each energy, apply the energy calibration to arrive at an average number of photons per frame, and convert this into an deposited energy within the sensor's integration time. For each data frame, we remove known hot pixels, then sum the image to yield $I$, the total signal measured in ADU. From $I$, we subtract a temperature-dependent background $B$ using the cubic spline fit derived in Sec. \ref{subsec:temp_cal} and the number of pixels contributing to this image. The total signal, $S = I - B$, is averaged over all 3000 frames, then divided by the centroid value $\mu(E)$ for the relevant X-ray energy predicted from the linear energy calibration (Sec. \ref{subsec:energy_cal}) to yield an average number of photons per frame. Finally, this number of photons per frame is converted to a detected power using the X-ray photon energy $E$ and the nominal exposure time $t =$ 135 ms. Explicitly, this is:

\begin{equation}
	\label{eq:power_imx290}
	P_{IMX290}(E) = \frac{\bar{S}}{(mE + b)} \frac{E}{t}.
\end{equation}

\begin{figure}
	\begin{center}
		\begin{tabular}{c}
			\includegraphics[width=5.5in]{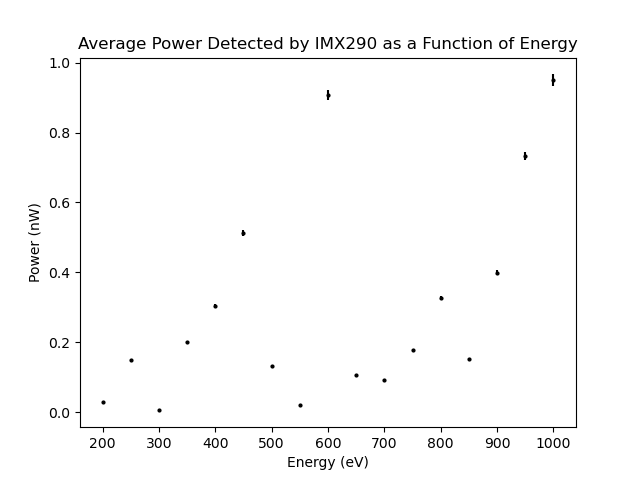}
		\end{tabular}
	\end{center}
	\caption 
	{ \label{fig:imx290-power}
		Power detected by the IMX290LLR-C as a function of energy, which ranges from approximately 5 -- 1000 pW (0.005 -- 1 nW). Errors are based on the quadrature sum of the error in the temperature-dependent background subtraction and the variance in the signal of the summed image, are typically  $\lesssim$1\% of the detected power, and appear smaller than the plot points.
		} 
\end{figure}

Resulting powers are shown in Fig. \ref{fig:imx290-power}. The uncertainty in $P_{IMX290}$ has contributions from errors in the energy calibration, variance in the total signal detected in each frame, and uncertainty in the photon energy and frame time. The IMX290LLR-C operates with a digital ``rolling shutter,'' in which the sensor is read out sequentially row-by-row. Pixels are reset before the next exposure, but are actively illuminated while the readout process is in progress. To account for this, we estimate an uncertainty in true exposure time equal to 2 ms, the readout time of the sensor under the chosen operating conditions. This is equivalent to a 1.5\% percent uncertainty, which is the dominant error in the measurement of the QE reported here. The errors in the fit parameters $m$ and $b$  the error in $b$ contribute $\lesssim$0.2\% each, while the uncertainty in the average total signal $\bar{S}$ is negligible ($\lesssim$0.05\%) as a result of averaging over 3000 data frames. Given the resolution of the beamline ($R \approx  7000$) implies an uncertainty in the X-ray energy of 0.02\%, we consider $E$ to be exact and exclude it from our error propagation.   

\section{Quantum Efficiency: Measurements \& Modeling}
\begin{figure}
	\begin{center}
		\begin{tabular}{c}
			\includegraphics[width=5.5in]{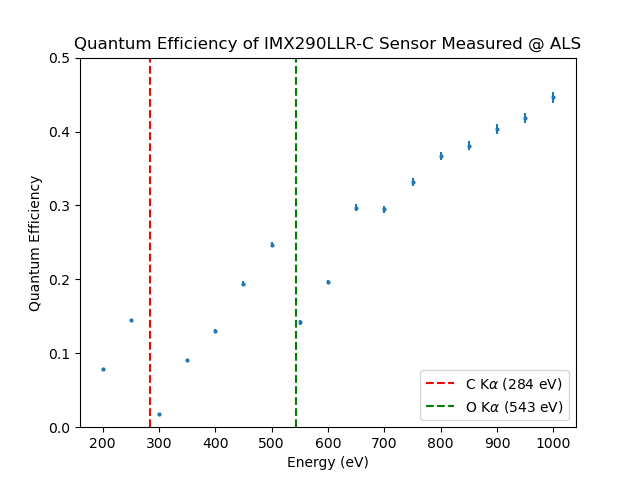}
		\end{tabular}
	\end{center}
	\caption 
	{ \label{fig:imx290-qe-raw}
		Quantum efficiency of the IMX290LLR-C as a function of soft X-ray energy.
		} 
\end{figure}

Since the photodiode and CMOS sensor both capture the whole beam profile, the QE of the IMX290LLR-C is the ratio between the detected power $P_{IMX290}(E)$ and the power in the direct beam $P_{Direct}(E)$, and is shown in Fig. \ref{fig:imx290-qe-raw}. The uncertainty associated with this QE measurement is straightforwardly propagated from estimated errors in $P_{Direct}(E)$ and $P_{IMX290}(E)$, shown in Figs. \ref{fig:photodiode-power} and \ref{fig:imx290-power} respectively, and is typically on the order of 2\% relative uncertainty. QE tends to increase as a function of X-ray energy, with two marked decreases just before the 300 eV and 550 eV data points. We flag the energies of the C K$\alpha$ (284 eV) and O K$\alpha$ (543 eV) absorption lines with dashed lines. 

\subsection{Microscopy of IMX290 Devices}
\label{subsec:microscopy}
\begin{figure}
	\begin{center}
		\begin{tabular}{c}
			\includegraphics[width=5.5in]{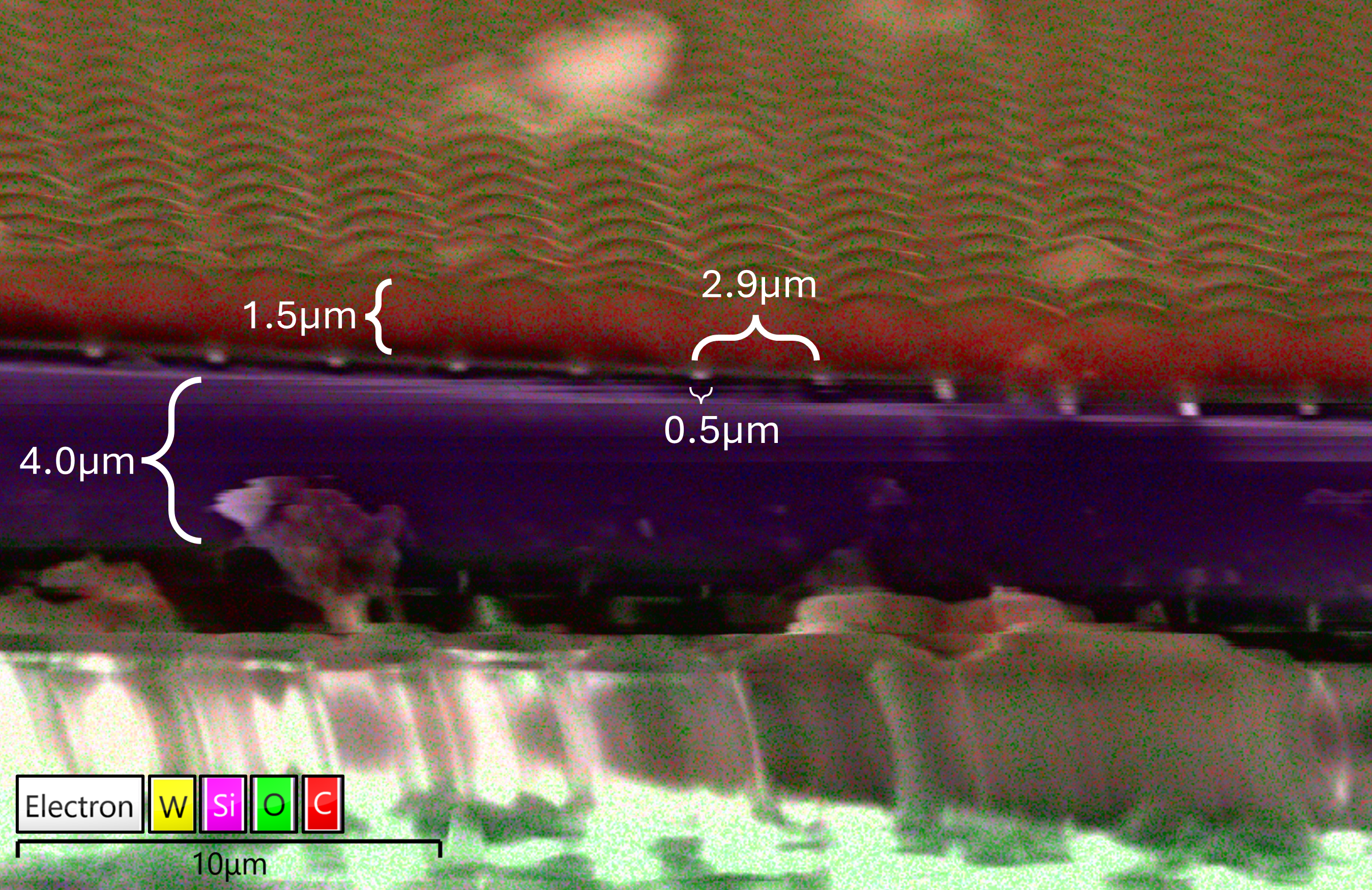}
		\end{tabular}
	\end{center}
	\caption 
	{ \label{fig:eds-sideon}
		An EDS micrograph of a cleaved, unaltered IMX290LLR-C device. Fluorescence from C, O, Si, and W are indicated by red, green, purple, and yellow coloration. To avoid contaminating the sample with unwanted elemental fluoresence, this semiconductor sample was not coated with a conductive layer. The bottom portion of this micrograph appears distorted, which we attribute to electron charging during the course of SEM measurements. 
		} 
\end{figure}

To inform a theoretical model of the IMX290LLR-C's QE, we investigated devices using optical and electron microscopy tools to understand possible layer elemental compositions and thicknesses.  As noted in Sec. \ref{subsec:experimental-setup}, the IMX290LLR-C device tested at Beamline 6.3.2 was subject to a mechanical scraping operation in an effort to remove the microlens array. To understand the layer compositions and thicknesses of an unmodified device (i.e., one not subject to mechanical scraping), we utilized a scanning electron microscope (SEM) equipped with an energy-dispersive X-ray spectroscopy (EDS) detector. A separate, unaltered IMX290LLR-C device was cleaved and mounted ``edge-on'' within the sample chamber of a JEOL JSM-1T700HR Scanning Electron Microscope. A layered EDS micrograph of this device is shown in Fig. \ref{fig:eds-sideon}. The image of elemental fluorescence lines of C $\alpha$ and O K$\alpha$ indicate that the microlens array is formed from photoresist capped with oxide. This array matches the device's pixel pitch of 2.9 $\upmu$m and measures approximately 1.5 $\upmu$m thick, while image formed from Si K$\alpha$ (1.7 keV) fluorescence shows a bulk Si layer measuring roughly 4.0 $\upmu$m. The Sony BI pixel architecture includes a metal aperture grid along the pixel boundaries to reduce crosstalk between the pixels under optical illumination \cite{Wakabayashi2010}. Our EDS analysis shows this structure and identifies this metal grid as W, in agreement with analysis performed on other Sony IMX chips \cite{Fontaine2015}.   



\begin{figure}
	\begin{center}
		\begin{tabular}{c}
			\includegraphics[width=5.5in]{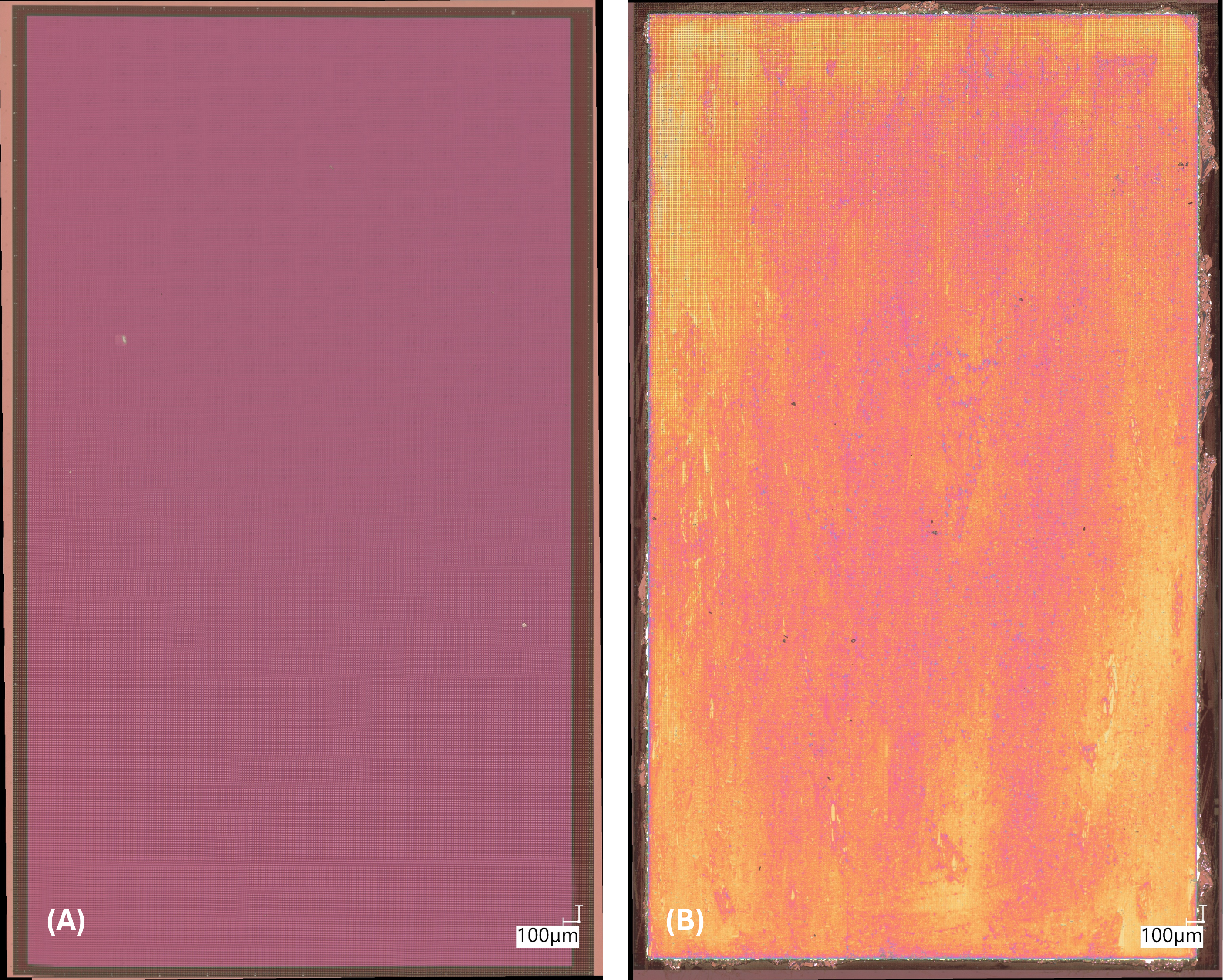}
		\end{tabular}
	\end{center}
	\caption 
	{ \label{fig:optical-microscope-comparison}
		A composite microscope image of the surface of a IMX290LLR-C device following (A) cover glass removal and (B) a mechanical scraping operation. A 100 $\upmu$m scale bar appears in the lower right of each image. The heterogenity of (B), along with the still visible microlens grid and purple coloration, indicate that a portion of the microlens array remains on the surface above the active Si.
		} 
\end{figure}

We next utilized a Keyence VHX-X1 digital microscope equipped with automated staging to examine the device surface. A high-magnification zoom lens (VH-Z250) compatible with the Keyence VHX-X1 was used to inspect the device at 1000$\times$ magnification, yielding a sampling of $\sim$0.26 $\upmu$m/pixel. A series of images covering the device area of 5.6 mm $\times$ 3.1 mm were autonomously recorded and stitched together in order to construct a map of the sensor surface. Fig. \ref{fig:optical-microscope-comparison}(A) shows the the composite image of a IMX290LLR-C without cover glass but otherwise unaltered, while Fig.\ref{fig:optical-microscope-comparison}(B) shows the device tested at ALS following mechanical scraping. With reference to Fig \ref{fig:optical-microscope-comparison}(B), we interpret the lack of uniformity across the sensor area, similarity in coloration to Fig. \ref{fig:optical-microscope-comparison}(A) over portions of the image, and the visible gridded structure remaining on this back-illuminated device as indicating incomplete removal of the microlens array via this method.

\subsection{Least-Squares Fitting of Measurement Data}
Armed with this information, we constructed a ``layer cake'' model of the IMX290LLR-C QE device and fit this model to our measurement data using the method of least squares. This model is of the form:

\begin{equation}
	\label{eq:qe_model}
	QE(E) = \mathcal{F} \left( \prod_{i=1}^{N} e^{-t_i/\lambda_i(E)} \right) \left(1 - e^{-t_{Si}/\lambda_{Si}(E)} \right)
\end{equation}

\noindent where $\mathcal{F}$ is a filling fraction of the detector capable of absorbing an X-ray photon, $t_i$ is the thickness of model layer $i$, and $\lambda_i(E)$ is the attenuation length within model layer $i$ at an energy $E$. The product series from $i$ = 0, ... $N$, models attenuation in $N$ layers above the active Si, while the remaining factor captures the probability of absorption in the active Si. We examine model components consisting of: (1) polymethyl methacrylate (PMMA), a common C-based photoresist, (2) diazonaphthoquinone (DNQ), a photoresist with a chemical composition including N, and (3) a dead layer of SiO$_{2}$. Attenuation depths $\lambda_{i}(E)$ for each layer are drawn from Center for X-ray Optics' database, which maintains an online interface to the Henke tables \cite{Henke1993}. Similar to Sec. \ref{subsec:energy_cal}, we employ \hyperlink{https://lmfit.github.io/lmfit-py/index.html}{\emph{lmfit}} to perform least-squares fitting of the model to the data. 

We note that the formulation of this model explicitly assumes that photons in the measured 200 - 1000 eV energy range incident on the W aperture grid are fully absorbed. This treatment of the W aperture grid is supported by noting that the attenuation depth of W is $\lesssim$ 130 nm across these energies. The $\gtrsim$250 nm W layer thickness implied by Fig. \ref{fig:eds-sideon} hence results in no more than $1/e^2 \sim 13.5\%$ transmission through this element at 1 keV, while the transmission at lower energies diminishes rapidly. As an alternative, we attempted a fit with a more complex model allowing for partial transmission through the W aperture grid. This approach fails to provide independent constraints on the tungsten layer thickness $t_W$ or $\mathcal{F}$, with a correlation coefficient (C$_{\mathcal{F}-t_{W}}$ $\gtrsim$ +0.9). This degeneracy implies that one of these parameters is unconstrained by these data, supporting the use of the simplified model.









\begin{figure}
	\begin{center}
		\begin{tabular}{c}
			\includegraphics[width=5.5in]{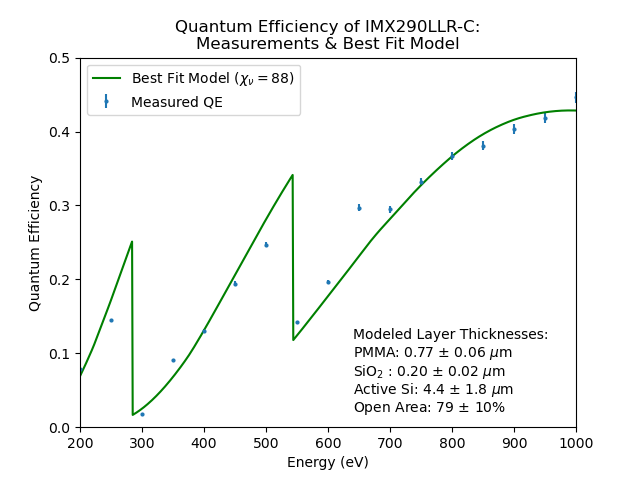}
		\end{tabular}
	\end{center}
	\caption 
	{ \label{fig:modeled_qe}
		The measured QE of the IMX290LLR-C plotted alongside with the best-fit ``layer cake'' model using a PMMA layer, SiO$_2$ passivation layer, active Si absorber, and a filling fraction. Derived errors are plotted with the data points, and in most cases are comparable to the marker size. 
		} 
\end{figure}

Our best fit model to the measured QE of the IMX290 is shown in Fig. \ref{fig:modeled_qe}.  The overall reduced chi-square $\chi_{\nu} = 88$ is well outside statistically acceptable values, suggesting that either our measurement errors are underestimated or that the employed model is too simplistic, with variations in parameters such as film chemical composition, stoichiometry, density, or uniformity that are not modeled. However, the derived model does provide an acceptable fit shape to the data from which the physics of the underlying detection can be roughly understood. The derived best fit model includes a filling fraction, SiO$_2$ and PMMA layers attenuating the X-ray flux, and an active Si layer. Using DNQ as a proxy for this family of resists, we find no evidence for photoresists incorporating nitrogen (best fit model w/ DNQ: $\chi_{\nu} = 222$). We also find that the inclusion of a filling factor $\mathcal{F}$ is strongly preferred, as a model in which $\mathcal{F}$ is fixed at unity yields both a degraded $\chi_{\nu} = 105$ and requires an active Si thickness $t_{Si} = 2.8 \pm 0.4  \upmu$m, a thickness inconsistent with EDS microscopy of the IMX290LLR-C.


The derived parameters from this best fit model compare favorably with the edge-on EDS measurements shown in Fig. \ref{fig:eds-sideon}. We find an active Si thickness of $t_{Si} = 4.4 \pm 1.8 \upmu$m, consistent with our EDS assessment of $\sim$4 $\upmu$m. Our derived filling fraction $\mathcal{F} = 0.79 \pm 0.10$ is consistent with interpreting the width of a grid element as measured in Fig. \ref{fig:eds-sideon} as blocking entrance into a pixel ($0.5\upmu$m / $2.9\upmu$m $\sim$ 20\%). The inferred photoresist thickness of $t_{PMMA} = 0.77 \pm 0.06 \upmu$m is below the thickness measured of an unaltered device, and consistent with our interpretation that the photoresist has been partially but incompletely removed. The derived SiO$_{2}$ thickness, $t_{SiO_{2}} = 0.20 \pm 0.02 \upmu$m, is too large for a native SiO$_{2}$ layer, which is typically on the order of a few nanometers. This derived layer parameter may represent a passivation layer atop this BI device or the effect of a residual SiO$_{2}$ capping layer atop a portion of the remaining microlens array. 

\section{Discussion and Conclusions}
Through an at-wavelength experiment at Beamline 6.3.2 of ALS, we have measured the QE of an IMX290LLR-C CMOS device prepared for detecting soft X-rays, finding an efficiency ranging from 0.1 -- 0.4 over the 200 -- 1000 eV bandpass. This series of measurements is in agreement with a previous experiment reporting the QE at a single energy of 490.5 eV of 28 $\pm$ 2\%\cite{Packard2024}.

From these QE measurements and supporting microscopy, we have developed a ``layer cake'' model of the IMX290LLR-C. This layer cake model suggests an active Si depth of approximately 4 $\upmu$m. Notably, we have found evidence that mechanical scraping does not result in complete removal of the microlens array from these IMX290LLR-C sensors. This operation has left a residual layer of photoresist as well as a grid supporting the microlens array which reduced the detection efficiency of the device via shadowing. 

Due to the presence of these attenuating layers and the relatively thin bulk Si, the efficiency of an IMX290LLR-C device falls well below that of other BI X-ray detectors employed for astronomy. For example, hybrid CMOS detectors achieve a quantum efficiency of 95\% at 1 keV \cite{Hossen2025}, while the back-illuminated CCDs aboard Suzaku achieve a QE of $\approx$90\% over a comparable energy range\cite{Koyama2007}. Perhaps unsurprisingly given the remaining material in the photon path before absorption in active Si, the IMX290LLR-C achieves a QE closer to the front-illuminated detectors aboard Suzaku and Chandra (10 -- 50\%)\cite{Koyama2007, Bautz1998}. In spite of the IMX290LLR-C's QE being below the present state-of-the-art, it may be still be useful for missions for which cost and system-level simplicity are major drivers, such as soft X-ray CubeSats or sounding rockets. The IMX290LLR-C is also useful in contexts where the reduced QE is an acceptable trade-off for small pixel sampling and excellent energy resolution, such as X-ray calibration facilities or synchrotron beamlines. The QE model presented here informs the sensitivity analyses of these applications.

Identifying an effective means of removing the microlens array from Sony's IMX class of sensors would further improve its X-ray QE and expand the range of its potential applications. The microlens array is aligned and added to the active Si sensor at a late stage in the sensor assembly process; however, manufacturing devices in which a microlens array is not added would require custom procurement, which may be infeasible within an industrial production setting and likely negates one of the chief advantages -- cost -- of the utilizing this commercially-available device. A more promising avenue for investigation is removal of the microlens array from fabricated devices; potential methods for doing so include chemical stripping \cite{Wilkes2016} or ablation via a high-powered laser \cite{Wright2023}. Following complete removal of this array, i.e., with only 200 nm thick SiO$_2$ passivation layer and 4.4 $\upmu$m of active Si remaining, the QE of the device improves to $\approx$70\% at 1 keV. Taken alongside investigations of achievable energy resolution and radiation tolerance, our results motivate continued investigations in adapting commercially-available CMOS sensors for use in X-ray astronomical missions.

\subsection*{Disclosures}
The authors declare that there are no financial interests, commercial affiliations, or other potential conflicts of interest that could have influenced the objectivity of this research or the writing of this paper.

\subsection* {Code, Data, and Materials Availability} 
The analysis scripts supporting this data are written in \emph{python} and developed within Jupyter Notebook. HTML exports readable by a standard web browser are available in a public-facing repository in the author's Github (https://github.com/ctderoo/imx290-qe). This manuscript analyzes thousands of full-frame images from an 2.1MPix IMX290LLR-C camera. This data totals 77.4 Gb, and hence cannot be hosted in the same repository. These data will be made available upon reasonable request directed to the corresponding author. 

\subsection* {Acknowledgments}
This effort was supported by NASA Grant 80NSSC24K0470 (``Technology Development for X-ray Small Satellite Missions'' : 23-RTF23-0001). This research used resources of the Advanced Light Source, a U.S. DOE Office of Science User Facility under contract no. DE-AC02-05CH11231. We are grateful for the test support provided by Eric Gullikson of Beamline 6.3.2. We acknowledge the University of Iowa Materials Analysis, Testing, and Fabrication Facility for use of the JEOL JSM-1T700HR Scanning Electron Microscope, which was purchased through funds provided by the National Science Foundation (Grant \#: MRI-2215495). We thank Mr. Michael Sinnwell and Mr. Schaffer Finney for their support with data collection and analysis. M. Weber acknowledges support from the Iowa Space Grant consortium's Undergraduate STEM Research Scholarship.


\bibliography{library}   
\bibliographystyle{spiejour}   


\vspace{2ex}\noindent\textbf{Casey DeRoo} received his BA degree in physics and classical studies in 2011 and his PhD in physics from University of Iowa in 2016. He worked at as an Astrophysicist at the Smithsonian Astrophysical Observatory prior to joining the faculty of the University of Iowa, where he is currently an Associate Professor. Prof. DeRoo was named a NASA Astrophysics Nancy Grace Roman Technology Fellow in 2020. His research interests include X-ray optics technologies, X-ray detectors, nanofabrication, and metrology.

\vspace{1ex}
\noindent Biographies of the other authors are not available.

\listoffigures
\listoftables

\end{spacing}
\end{document}